\documentclass[twocolumn,prb,showpacs]{revtex4-1}
\usepackage{graphicx}
\usepackage{ulem}
\newcommand{\ds}{\displaystyle}
\newcommand{\up}{\uparrow}
\newcommand{\dn}{\downarrow}
\newcommand{\de}{\partial}
\newcommand{\xc}{\rm xc}

\begin{document}

\title{
Response properties of III-V dilute magnetic semiconductors: interplay of disorder, dynamical electron-electron interactions
and band-structure effects}

\author{F. V. Kyrychenko}
\author{C. A. Ullrich}
\affiliation{Department of Physics and Astronomy, University of
Missouri, Columbia, Missouri 65211, USA}

\date{\today}

\begin{abstract}
A theory of the electronic response in spin and charge disordered media is developed with the particular aim to describe III-V dilute
magnetic semiconductors like Ga$_{1-x}$Mn$_x$As. The theory combines a detailed ${\bf k\cdot p}$ description of the valence band,
in which the itinerant carriers are assumed to reside, with first-principles calculations of disorder
contributions using an equation-of-motion approach for the current response function.
A fully dynamic treatment of electron-electron interaction is achieved by means of time-dependent density functional
theory. It is found that collective excitations within the valence band significantly increase the carrier relaxation rate
by providing effective channels for momentum relaxation.
This modification of the relaxation rate, however, only has a minor impact on
the infrared optical conductivity in Ga$_{1-x}$Mn$_x$As, which is mostly determined by the details of the valence band structure and
found to be in agreement with experiment.
\end{abstract}

\pacs{72.80Ey, 75.50Pp, 78.20.Bh} \keywords{dilute magnetic semiconductors,
resistivity, scattering}

\maketitle

\section{Introduction}
The idea of using both charge and spin of electrons in a new generation of electronic devices constitutes
the basis of spintronics. \cite{spintronic} The magnetic properties of the material
combined with its semiconducting nature makes dilute magnetic semiconductors (DMSs) potentially appealing
for various spintronics applications. \cite{ohno} In particular, the effect of carrier mediated ferromagnetism
opens up the possibility to control the electron spin and magnetic state of a system or device by means of an electric
field. A lot of attention is drawn to Ga$_{1-x}$Mn$_x$As due to the well developed technology of the
conventional GaAs based electronics and discovery of its relatively high ferromagnetic transition
temperature, \cite{ohno} with a current record of $T_c = 185$ K. \cite{record}

Unlike most other III-V DMSs, the nature of the itinerant carriers in Ga$_{1-x}$Mn$_x$As is still under
debate.\cite{jungwirth_review,burch_review} It is widely accepted that for low-doped insulating samples
the Fermi energy lies in a narrow impurity band. For more heavily doped, high-$T_c$ metallic samples there
are strong indications that the impurity band merges with the host semiconductor valence band forming
mostly host-like states at the Fermi energy with some low-energy tail of disorder-related localized
states. \cite{jungwirth07PRBtails} First-principles calculations \cite{mahadevan,sandratskii,yildrim}
have so far not been fully conclusive regarding the nature of the itinerant carriers in this regime,
and further theoretical studies continue to be necessary. Meanwhile, attention has shifted to model
Hamiltonian approaches assuming either the valence band \cite{dietl01} or impurity band \cite{berciu}
picture and their ability to adequately describe the experimental results in Ga$_{1-x}$Mn$_x$As.

Most theoretical approaches assuming the valence-band nature of itinerant holes in Ga$_{1-x}$Mn$_x$As treat the
band structure in detail, while disorder and many-body effects are only accounted for using simple
phenomenological relaxation time approximations and static screening models. \cite{sinova,rates,ewelina}
On the other hand, the extreme sensitivity of magnetic and transport properties of Ga$_{1-x}$Mn$_x$As
to details of the growth conditions \cite{shimizu} and post-growth annealing \cite{hayashi,potashnik,yu}
points to the crucial role played by the defects and their configurations, and has stimulated intense
research on the structure of defects and their influence on the various properties of the system. \cite{timm_rev}
It is essential, therefore, to develop a theory of electrical conductivity in DMSs with more emphasis
given to disorder and electron-electron interactions, without neglecting the intricacies of the electronic band structure.

Here we present a comprehensive theory for the electron dynamics in DMSs which accounts for the complexity
of the valence band structure of the semiconductor host material and treats disorder and electron-electron
interaction on an equal footing. In previous work we used a simplified treatment of the semiconductor
valence band \cite{KUPRB07,KUjpcm09} or considered only static properties of the system. \cite{KUPRB09}
In this paper we simultaneously account for the complexity of the valence band, use a first-principles
approach to describe disorder contributions, and employ a fully dynamic treatment of electron interactions.

To account for the valence band structure we use the generalized ${\bf k\cdot p}$ approach\cite{birpikus}
where a certain number of bands are treated exactly while the contribution from the remote bands is
included up to second order in momentum. To describe disorder effects we use the
equation of motion for the paramagnetic current response function of the fully disordered system.
This approach has some similarities to models developed earlier using the memory function
formalism. \cite{Gotze81,belitz,UV} The advantage of our approach as compared to the memory function
formalism is the relative simplicity and transparency of the derivation and the straightforward
possibility to include the spin degree of freedom. Another advantage is that our formalism is
expressed in terms of a current-current and a set of density and spin-density response functions.
This enables us to use the powerful apparatus of time-dependent density-functional theory (TDDFT) \cite{TDDFT}
to treat many-body effects such as dynamic screening and collective excitations of the itinerant carriers in principle exactly.

The paper is divided into two major sections and conclusions. For ease of reading, some of the derivations are presented
in appendices. The theory section (Sec.~\ref{sec:theory}) is organized as follows. In Sec.~\ref{sec:general}
we present our general formalism based on the equation of motion of the current-current response
function of the disordered system. In Sec.~\ref{sec:kp} we describe the evaluation of the current-current,
density and spin-density response functions for the multiband system using a generalized ${\bf k\cdot p}$
perturbation approach. Next, in Sec.~\ref{sec:e-e} we show the treatment of electron-electron interaction
by means of TDDFT. In Section \ref{sec:results} we first discuss the new
features that the valence band character of itinerant carriers brings into the system, namely the
dominance of the long-wavelength side of the single-particle excitation spectrum by the interband
spin transitions and the effective suppression of the collective plasmon excitations within the
valence band for the whole range of momentum. Next, in Sec.~\ref{sec:doped} we discuss the effect
of magnetic doping: spin and charge disorder in the system and modification of the band structure
in the magnetically ordered phase. We show that the full dynamic treatment of electron-electron interactions
allows us to capture the effect of collective excitations on the carrier relaxation time. We then compare
our results also with experimental data on infrared conductivity. Finally, in Sec.~\ref{sec:conclusion}
we draw our conclusions.

\section{Theory}\label{sec:theory}

\subsection{General formalism}\label{sec:general}

We discuss a system described by the Hamiltonian
\begin{equation}\label{H}
  \hat{H}=\hat{H}_e+\hat{H}_m+\hat{H}_d,
\end{equation}
where $\hat{H}_e$ is the contribution of the itinerant carriers and $\hat{H}_m$ represents
the subsystem of localized magnetic spins.
These two terms constitute the ``clean''  part of the total Hamiltonian. The last term in Eq.~(\ref{H})
describes disorder in the system:
\begin{equation}\label{HI4}
  \hat{H}_d = V^2 \sum_{\bf k} \hat{\vec{\cal{U}}}({\bf k})\cdot \hat{\vec{\rho}}({\bf -k}),
\end{equation}
where the four-component charge and spin disorder scattering potential
\begin{equation}\label{pot}
  \hat{\vec{\cal{U}}}({\bf k})=\frac1{V}\sum_j \left(\begin{array}{c}
    U_j({\bf k}) \\
    -\frac{J}2\left( \hat{S}_j^z-\langle S \rangle \right) \\
    -\frac{J}2 \hat{S}_j^- \\
    -\frac{J}2 \hat{S}_j^+ \
  \end{array} \right) e^{i{\bf k\cdot R}_j}
\end{equation}
is coupled to the four-component vector of charge and spin density operators of the itinerant carriers:
\begin{equation}\label{rho4}
  \hat{\vec{\rho}}=\left(\begin{array}{l}
    \hat{\rho}^1 \\
    \hat{\rho}^z \\
    \hat{\rho}^+ \\
    \hat{\rho}^- \
  \end{array}\right)=\left(\begin{array}{l}
    \hat{n} \\
    \hat{s}^z \\
    \hat{s}^+ \\
    \hat{s}^- \
  \end{array} \right)
\end{equation}
with the components
\begin{equation}\label{rho}
  \hat{\rho}^{\mu}({\bf k})=\frac1{V}\sum_{\bf q} \sum_{n n'} \langle u_{n',{\bf q-k}}|\sigma^{\mu}| u_{n,{\bf q}}\rangle \;
  \hat{a}^+_{n',{\bf q-k}} \, \hat{a}_{n,{\bf q}} \:.
\end{equation}
Here, $\sigma^\mu$ ($\mu=1,z,+,-$) is defined via the Pauli
matrices, where $\sigma^1$ is the $2\times 2$ unit matrix,
$\sigma^\pm=(\sigma^x \pm i \sigma^y)/2$, and $| u_{n,{\bf
q}}\rangle$ are the two-component Bloch function spinors with wave
vector ${\bf q}$ and band index $n$. The summation in
Eq.~(\ref{pot}) is performed over all defects. Note that the mean field part of the $p$-$d$ exchange interaction between
itinerant holes and localized spins is absorbed into the clean system band structure Hamiltonian $\hat{H}_e$;
disorder in our model consists of the Coulomb potential of charge defects and fluctuations of localized
spins around the mean field value $\langle S \rangle$.

The general case of multiple types of defects, including defect
correlations, was considered in Ref.~\onlinecite{KUPRB07}. For
simplicity we here include only the most important defect type,
namely randomly distributed manganese ions in gallium substitutional
positions (${\rm Mn_{Ga}}$). Our model treats localized spins as
quantum mechanical operators coupled to the band carriers via a
contact Heisenberg interaction featuring a momentum-independent
exchange constant $J$. We use the value of $VJ=-55\:{\rm meV\,nm^3}$,
which corresponds to the widely used DMS $p$-$d$ exchange constant
$N_0 \beta=-1.2\,$eV. \cite{dietl01} The $z$-axis is chosen along the
direction of the macroscopic magnetization.

Earlier we  developed a theory of transport in charge and spin disordered media with emphasis on
a treatment of disorder and electron-electron interaction.\cite{KUjpcm09}  It is based on an equation
of motion \cite{Gotze72,GV} approach for the paramagnetic current-current response of the full, disordered system:
\begin{equation}\label{chipar}
  \chi_{j_{p\alpha} j_{p\beta}}({\bf r},{\bf r'},\tau)=-\frac{i}{\hbar}\Theta(\tau)
  \langle [\hat{j}_{p\alpha}(\tau,{\bf r}),\hat{j}_{p\beta}({\bf r'})]\rangle_{H},
\end{equation}
where
\begin{equation}
  \hat{j}_{p\alpha}(\tau,{\bf r})=
  e^{\frac{i}{\hbar}\hat{H}\tau}\hat{j}_{p\alpha}({\bf r})e^{-\frac{i}{\hbar}\hat{H}\tau}
\end{equation}
is the paramagnetic current-density operator in Heisenberg representation
and $\alpha,\beta=x,y,z$ are Cartesian coordinates. During the derivation we assumed our system to be
macroscopically homogeneous, which is justified if the coherence length of the electrons is much shorter
than the system size. In this case, summing over
all electrons will leave us with an averaged effect of disorder that does not depend on the particular
disorder configuration. For such macroscopically homogeneous systems the response at point ${\bf r}$
depends only on the distance $|{\bf r-r'}|$ to the perturbation and not on the particular
choice of points ${\bf r}$ and ${\bf r'}$.  Another major approximation involves the decoupling procedure,
where we neglect the influence of the itinerant carriers on the localized spins.
Therefore, our approach does not include magnetic polaron effects and lacks the microscopic features
of carrier mediated ferromagnetism. The latter, however, can be
reinstated to some extent by introducing a phenomenological Heisenberg-like term in the magnetic subsystem
Hamiltonian $\hat{H}_m$. Details of the derivation are presented in Ref. \onlinecite{KUjpcm09}.

The final expression for the total current response reads
\begin{eqnarray}\label{finalS}
\chi^J_{\alpha \beta}({\bf q},\omega) &=& \chi^c_{j_{p\alpha} j_{p\beta}}({\bf q},\omega) + \frac{n}{m}\delta_{\alpha \beta}
\nonumber\\
&+&
\frac{V^2}{m^2 \omega^2}\sum_{\bf k} k_{\alpha} k_{\beta}
\sum_{\mu \nu} \left\langle\hat{\cal{U}}_{\mu}({\bf k})\; \hat{\cal{U}}_{\nu}({\bf -k})\right\rangle_{H_m}
\nonumber\\
&&
\times  \Big(\chi_{\rho^{\mu}\rho^{\nu}}({\bf q-k},\omega)-\chi^c_{\rho^{\mu} \rho^{\nu}}({\bf -k}) \Big),
\end{eqnarray}
where $\chi_{\rho^{\mu}\rho^{\nu}}({\bf k},\omega)$ is the set of charge and spin density response functions
with respect to the operators (\ref{rho4})--(\ref{rho}) and the superscript ``$c$'' indicates quantities defined in the clean system.
By comparing Eq.~(\ref{finalS}) with the Drude formula in the weak disorder limit $\omega \tau \gg 1$,
\begin{equation}
  \chi^J_D(\omega)=\frac{n}{m}\frac1{1+i/\omega\tau}\approx \frac{n}{m}-\frac{i n}{m \omega \tau},
\end{equation}
we identify the tensor of Drude-like frequency- and momentum-dependent relaxation rates of the form
\begin{eqnarray}\label{tau}
\tau^{-1}_{\alpha\beta}({\bf q},\omega)  & = & i \frac{V^2}{n m
\omega}\sum_{ \bf k \atop \mu\nu } k_{\alpha} k_{\beta}
  \left\langle\hat{\cal{U}}_{\mu}({\bf -k})\;
  \hat{\cal{U}}_{\nu}({\bf k})\right\rangle_{H_m}  \nonumber\\
&\times&
  \Big(\chi_{\rho^{\mu}\rho^{\nu}}({\bf q-k},\omega)-\chi^c_{\rho^{\mu} \rho^{\nu}}({\bf k},0) \Big).
\end{eqnarray}

Note that the right-hand side of Eqs.~(\ref{finalS}) and (\ref{tau}) contains the set of spin and
charge response functions of the full, disordered system. Therefore, strictly speaking, Eq.~(\ref{finalS})
should be evaluated self-consistently \cite{Gold} with the continuity equations closing the loop.
Here we use a simplified approach based on two approximations. First, taking the weak disorder limit
in the right hand side of Eq.~(\ref{tau}) we retain terms up to the second order in components of the
disorder potential. In other words, the spin and charge response functions of the full system in Eq.~(\ref{tau})
are replaced by their clean system counterparts:
\begin{equation}
 \chi_{\rho^{\mu}\rho^{\nu}}({\bf q-k},\omega) \to \chi^c_{\rho^{\mu}\rho^{\nu}}({\bf q-k},\omega).
\end{equation}
Next we assume that the paramagnetic current response function of the full system may be expressed
as the clean system response function with a lifetime broadening given by Eq.~(\ref{tau}):
\begin{equation}\label{self}
 \chi_{j_{p\alpha} j_{p\beta}}({\bf q},\omega) \approx \chi^c_{j_{p\alpha} j_{p\beta}}({\bf q},\omega-i\tau^{-1}_{\alpha\beta}).
\end{equation}
Equations (\ref{tau})--(\ref{self}) will be used in the following.

\subsection{Multiband ${\bf k \cdot p}$ approach} \label{sec:kp}

In order to obtain the conductivity through Eqs.~(\ref{tau})--(\ref{self}) we will have to calculate
the paramagnetic current response and spin and charge density response functions of the clean system.
To properly describe the complexity of the semiconductor valence band we are going to implement the
multiband ${\bf k\cdot p}$ approach.

First we derive the current and density response functions in the formal basis of the Bloch states
\begin{equation}\label{Bbasis}
  |n,{\bf k}\rangle=\frac1{\sqrt{V}} e^{i{\bf k\cdot r}} |u_{n,{\bf k}} \rangle
\end{equation}
which diagonalize the clean system Hamiltonian
\begin{equation}
  \hat{H}=\sum_{n,{\bf k}} \varepsilon_{n,{\bf k}} \; \hat{a}^+_{n,{\bf k}} \hat{a}_{n,{\bf k}}.
\end{equation}
Within second quantization in the basis (\ref{Bbasis}), the paramagnetic current in the system with spin-orbit interaction is given by

\begin{eqnarray} \label{jpM}
\hat{\bf j}_p({\bf q})
&=&
\frac1{V} \sum_{n,n',{\bf k}} \left[\frac{\hbar}{m_0} \left({\bf k}-\frac12 {\bf q}\right)
  \langle u_{n',{\bf k-q}} | u_{n,{\bf k}}\rangle \right.
\nonumber\\
&+&
\left.\frac1{m_0}\langle u_{n',{\bf k-q}} | \hat{\vec{\pi}} | u_{n,{\bf k}}\rangle \right]
\hat{a}^+_{n',{\bf k-q}} \hat{a}_{n,{\bf k}},
\end{eqnarray}
with
\begin{equation}\label{pi}
  \hat{\vec{\pi}}=\hat{\bf p}+\frac{\hbar}{4 m_0 c^2} [\hat{\bf \sigma}\times \hat{\nabla} U_c],
\end{equation}
where $U_c$ is the periodic crystal field potential.
Hereafter, performing the real space integration we assume that the envelop function varies slowly on the scale of the unit cell.

Introducing the time dependence of the creation and destruction operators in (\ref{jpM}),
the paramagnetic current response of the multiband system can be directly evaluated, and one finds
\begin{widetext}
\begin{eqnarray}\label{currentMB}
  \chi^c_{j_{p\alpha} j_{p\beta}}({\bf q},\omega) &=& \frac{1}{V m_0^2} \sum_{n,n',{\bf k}}
  \frac{f_{n',{\bf k-q}}-f_{n,{\bf k}}}{\varepsilon_{n',{\bf k-q}}-\varepsilon_{n,{\bf k}}+\hbar\omega +i \eta} \\
  &\times&\left[\hbar\left(k_{\alpha}-\frac{q_{\alpha}}2\right) \langle u_{n',{\bf k-q}}  | u_{n,{\bf k}}\rangle+
  \langle u_{n',{\bf k-q}} |\hat{\pi}_{\alpha} | u_{n,{\bf k}}\rangle \right]
  \left[\hbar\left(k_{\beta}-\frac{q_{\beta}}2\right) \langle u_{n,{\bf k}}  | u_{n',{\bf k-q}}\rangle+
  \langle u_{n,{\bf k}} |\hat{\pi}_{\beta} | u_{n',{\bf k-q}}\rangle \right]. \nonumber
\end{eqnarray}
A similar procedure for the spin and charge density response yields:
\begin{equation}\label{spindensityMB}
  \chi^c_{\rho^{\mu} \rho^{\nu}}({\bf q},\omega) = \frac{1}{V} \sum_{n,n',{\bf k}}
  \frac{f_{n',{\bf k-q}}-f_{n,{\bf k}}}{\varepsilon_{n',{\bf k-q}}-\varepsilon_{n,{\bf k}}+\hbar\omega +i \eta}
  \; \langle u_{n',{\bf k-q}} |\hat{\sigma}^{\mu} | u_{n,{\bf k}}\rangle
  \langle u_{n,{\bf k}} |\hat{\sigma}^{\nu} | u_{n',{\bf k-q}}\rangle.
\end{equation}

All we need now for evaluating Eqs.~(\ref{currentMB}) and (\ref{spindensityMB}) is to determine the
form of the periodic Bloch functions $|u_{n,{\bf k}} \rangle$ that diagonalize the clean system Hamiltonian.
The common approach is to diagonalize the multiband ${\bf k\cdot p}$ Hamiltonian that treats
certain bands exactly and treats contributions from remote bands up to second order in momentum.
The derivation of such a Hamiltonian is outlined in Appendix~\ref{app:kp}. By diagonalizing the
 matrix of this Hamiltonian, however, we obtain the eigenvectors of the modified Hamiltonian (\ref{Hbar}).
 Before evaluating the matrix elements between Bloch periodic functions  $|u_{n,{\bf k}} \rangle$ in
 Eqs.~(\ref{currentMB}) and (\ref{spindensityMB}) we therefore have to perform the unitary transformation (\ref{canon}).
 Details of these calculations are presented in Appendix~\ref{app:me}.

The final expression for the paramagnetic current response function in the long-wave limit ${\bf q}=0$
(since we are looking for the optical response) is given by
\begin{eqnarray}
  \chi^c_{j_{p\alpha} j_{p\beta}}(\omega) &=& \frac{1}{V m_0^2} \sum_{n,n',{\bf k}}
  \frac{f_{n',{\bf k}}-f_{n,{\bf k}}}{\varepsilon_{n',{\bf k}}-\varepsilon_{n,{\bf k}}+\hbar\omega +i \eta} \\
  &\times&\left[\sum_{s' s} B^*_{s'}(n',{\bf k}) B_s(n,{\bf k})\frac{m_0}{\hbar} \frac{\de}{\de k_{\alpha}} \langle s' |\bar{H}| s \rangle \right]
  \left[\sum_{s' s} B^*_{s}(n,{\bf k}) B_{s'}(n',{\bf k})\frac{m_0}{\hbar} \frac{\de}{\de k_{\beta}} \langle s |\bar{H}| s' \rangle  \right], \nonumber
\end{eqnarray}
where $\bar{H}$ denotes the effective multiband ${\bf k\cdot p}$ Hamiltonian (\ref{Hbar})
and ${\bf B}(n,{\bf k})$ is its eigenvector for the state with energy $\varepsilon_{n,{\bf k}}$.
The charge and spin density response is approximated by
\begin{eqnarray}\label{denq}
  \chi^c_{\rho^{\mu} \rho^{\nu}}({\bf q},\omega) & \approx & \frac{1}{V} \sum_{n,n',{\bf k}}
  \frac{f_{n',{\bf k-q}}-f_{n,{\bf k}}}{\varepsilon_{n',{\bf k-q}}-\varepsilon_{n,{\bf k}}+\hbar\omega +i \eta} \\ \nonumber
  &\times& \sum_{s', s, \tau, \tau'} B^*_{s'}(n',{\bf k-q}) B_{\tau'}(n',{\bf k-q}) B_s(n,{\bf k}) B^*_{\tau}(n,{\bf k})
  \langle s' |\hat{\sigma}^{\mu}|s\rangle \langle \tau |\hat{\sigma}^{\nu}|\tau' \rangle.
\end{eqnarray}
If $\hat{\sigma}^{\mu}=\left(\hat{\sigma}^{\nu}\right)^+$, i.e. for $\chi_{nn}$, $\chi_{s^z s^z}$ and $\chi_{s^{\pm}
s^{\mp}}$, the second sum is a real quantity. Then, the imaginary part is
\begin{equation}\label{imchi}
   \Im[\chi^c_{\rho^{\mu} (\rho^{\mu})^+}({\bf q},\omega)] = -\frac{\pi}{(2\pi)^3} \sum_{n,n'}
  \int d^3 {\bf k} (f_{n',{\bf k-q}}-f_{n,{\bf k}}) \delta[\hbar\omega-(\varepsilon_{n,{\bf k}}-\varepsilon_{n',{\bf k-q}})]
  \left| \sum_{s', s'} B^*_{s'}(n',{\bf k-q}) B_s(n,{\bf k}) \langle s' |\hat{\sigma}^{\mu}|s\rangle \right|^2.
\end{equation}
\end{widetext}
It is seen that in the long-wavelength limit (${\bf q}\to 0$) the imaginary part of the density response
($\sigma^{\mu}\equiv \sigma^1$) vanishes as a product of orthogonal states, while the imaginary part of
spin response is, in general, finite. We conclude from this that the long-wavelength spectrum of
single-particle excitations is dominated by spin transitions.

The calculations were performed within an 8-band ${\bf k\cdot p}$ model. The basis functions and explicit
form of the Hamiltonian matrix are presented in Appendix~\ref{app:8band}.

\subsection{Electron-electron interaction}\label{sec:e-e}

A major advantage of our formalism is that it is expressed in terms of current and density response functions.
This allows us to use the powerful apparatus of TDDFT to account for the effects
of electron-electron interaction.

Let us first examine the current response of the clean system. In this
paper we are considering the optical response, i.e. the response to transverse perturbations.
Since transverse perturbations only induce a transverse response in a homogeneous system, there are
no density fluctuations directly created by an electromagnetic field. The total current response of
the interacting system in this case can be expressed as
\begin{equation}
  \Big(\chi^J({\bf q},\omega)\Big)^{-1}=
  \Big(\chi_0^J({\bf q},\omega)\Big)^{-1}+\frac{4 \pi e}{\omega^2-c^2q^2}+
  \frac{q^2}{\omega^2}v_q G_{T+},
\end{equation}
where $\chi_0^J$ is the response of the noninteracting system, $v_q$ is the Coulomb interaction, and the local
field factor $G_{T+}$ represents corrections from the exchange-correlation (xc) part of the electron interaction.

The corrections to the transverse current response function caused by electron-electron interaction are
relativistically small in this case and can be neglected. So, for the transverse current response of the
clean system we will use the noninteracting form.

The set of the density and spin-density response functions of the clean system enters our expression
(\ref{tau}) for the frequency- and momentum-dependent relaxation rates.
TDDFT allows us to describe all the effect of electron interaction, including
correlations and collective modes, in principle, exactly. Within the TDDFT formalism the charge- and
spin-density responses of the interacting system can be expressed
as:\cite{grosskohn}
\begin{equation}\label{tddft}
  \uuline{\chi}^{-1}({\bf q},\omega)=\uuline{\chi_0}^{-1}({\bf q},\omega) - \uuline{v}(q) - \uuline{f_{\xc}} ({\bf q},\omega),
\end{equation}
where all quantities are $4\times 4$ matrices and $\uuline{\chi_0}$ denotes the matrix of response
functions of the noninteracting system, $\uuline{v}(q)$ is the Hartree part of the electron-electron
interactions, and $\uuline{f_{\xc}}$ represents xc corrections in the form of local
field factors. As a simplification we use only the exchange part of $\uuline{f_{\xc}}$ and apply the adiabatic
local spin density approximation. Explicit expressions for the local field factors of the partially spin
polarized system are given in Appendix~\ref{app:lff}.

In general, $\uuline{f_{\xc}}$ is a symmetric $4\times 4$ matrix. If, however, the $z$-axis is directed
along the average spin, then the ground state transversal spin densities vanish,  $\rho_+=\rho_-=0$, and the matrix $\uuline{f_{\xc}}$
becomes block-diagonal:
\begin{equation}\label{fxc}
  \uuline{f_{\xc}}=\left(
  \begin{array}{cccc}
    f_{11} & f_{1z} & 0 & 0 \\
    f_{1z} & f_{zz} & 0 & 0 \\
    0 & 0 & 0 & f_{+-} \\
    0 & 0 & f_{+-} & 0 \
  \end{array}\right).
\end{equation}
Performing the matrix inversion in Eq.~(\ref{tddft}) we obtain the tensor of response functions of the interacting system in the form
\begin{widetext}
\begin{equation}
  \uuline{\chi}\equiv \left(
  \begin{array}{cccc}
   \chi_{nn} & \chi_{ns^z} & \chi_{ns^+} & \chi_{ns^-} \vspace{4mm} \\ \vspace{4mm}
   \chi_{s^zn} & \chi_{s^z s^z} & \chi_{s^z s^+} & \chi_{s^z s^-} \\ \vspace{4mm}
   \chi_{s^+n} & \chi_{s^+ s^z} & \chi_{s^+ s^+} & \chi_{s^+ s^-} \\ \vspace{4mm}
   \chi_{s^-n} & \chi_{s^- s^z} & \chi_{s^- s^+} & \chi_{s^- s^-}
  \end{array}\right)=
\left(
  \begin{array}{cccc}
    \ds \frac{\chi_{nn}^0-f_{zz}\Delta}{\varepsilon_{\rm LFF}} \vspace{3mm} &
    \ds \frac{\chi_{ns^z}^0+f_{1z}\Delta}{\varepsilon_{\rm LFF}} & 0 & 0 \\
    \ds \frac{\chi_{s^z n}^0+f_{1z}\Delta}{\varepsilon_{\rm LFF}} \vspace{3mm} &
    \ds \frac{\chi_{s^z s^z}^0-(v(q)+f_{11})\Delta}{\varepsilon_{\rm LFF}} & 0 & 0 \\
    0 & 0 & 0 &\ds \frac{\chi_{s^+s^-}^0}{1-f_{+-}\chi_{s^+s^-}^0} \\
    0 & 0 & \ds \frac{\chi_{s^-s^+}^0}{1-f_{+-}\chi_{s^-s^+}^0}& 0 \
  \end{array}\right),
\end{equation}
where
\begin{equation}
  \varepsilon_{\rm LFF}=1-\big(v(q)+f_{11}\big)\chi_{nn}^0(q,\omega)-f_{zz}\chi_{s^z s^z}^0(q,\omega)-
  f_{1z} \left(\chi_{n s^z}^0(q,\omega)+\chi_{s^z n}^0(q,\omega)\right)+
  \Big(f_{zz}\big(v(q)+f_{11}\big)-f_{1z}^2\Big)\Delta,
\end{equation}
\end{widetext}
and
\begin{equation}
  \Delta=\chi_{nn}^0 \chi_{s^z s^z}^0-\chi_{n s^z}^0\chi_{s^z n}^0=4\chi_{\up}^0\chi_{\dn}^0.
\end{equation}

\section{Results and discussion}\label{sec:results}

\begin{figure}
\centering
\includegraphics[width=1.0\linewidth]{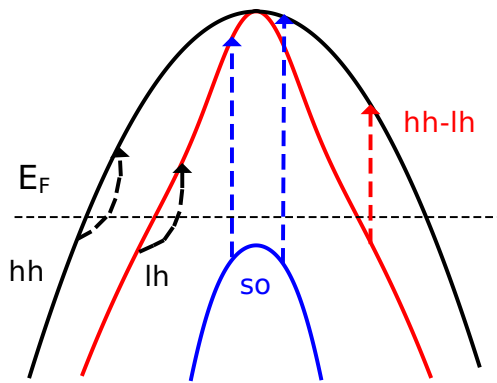}
\caption{(Color online) Schematic diagram of the possible single-particle excitations in the valence band
of a $p$-type semiconductor. Dashed lines indicate intra-valence band excitations within the heavy hole band (hh),
within the light hole band (lh) and inter-valence band excitations between heave hole and light hole bands (hh-lh)
and between split-off and heavy hole and light hole bands (so).} \label{fig1}
\end{figure}

We now discuss applications of our formalism for the specific case of GaMnAs DMSs. The band structure
parameters used in our calculations correspond to those of the GaAs host material: the band gap and spin-orbit
splitting are $E_g=1.519$ eV and $\Delta=0.341$ eV, Luttinger parameters are $\gamma_1=6.97$, $\gamma_2=2.25$ and
$\gamma_3=2.85$, conduction band effective mass is $m_e=0.065\; m_0$, Kane momentum matrix element is $E_p=27.86$ eV
and the static dielectric constant $K=13$. The $s(p)$-$d$ exchange interaction constants within the
conduction and valence bands are $N_0 \alpha=0.2$ eV and $N_0\beta=-1.2$ eV, respectively.

\subsection{Clean p-type GaAs}\label{sec:clean}

Before considering the effects of magnetic impurities and associated charge and spin disorder on
the transport properties, we would like to discuss some new features that the valence band character of
the itinerant carriers brings into the system. They stem from the complexity of the semiconductor valence band:
strong spin orbit interaction and the $\Gamma$-point degeneracy of the $p$-states.

\begin{figure}[t]
\centering
\includegraphics[width=0.95\linewidth]{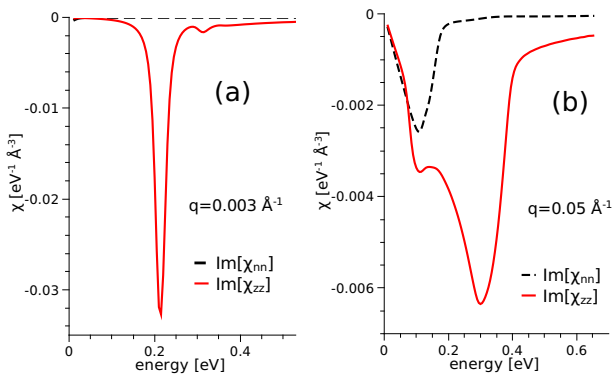}
\caption{(Color online) Imaginary part of the noninteracting density and longitudinal spin response functions
in $p$-doped GaAs for different wave vectors $q=0.003\mbox{\AA}^{-1}$ (a) and $q=0.05\mbox{\AA}^{-1}$ (b). The hole
concentration is $p=3.5\times 10^{20} \; {\rm cm}^{-3}$.} \label{fig2}
\end{figure}

The multiband nature of the valence band gives rise to a rich single-particle excitation spectrum.
In Fig.~\ref{fig1} we show a schematic representation of the valence band structure of a $p$-type semiconductor.
Arrows indicate the possible single-particle excitations. In addition to the intraband excitations within
the heavy hole band (analogous to the excitations within the conduction band of $n$-doped semiconductors),
here we have intra-band excitations within the light hole band as well as inter-valence band excitations
between light and heavy hole bands and between split-off and heavy and/or light hole bands.

The variety of the possible single-particle excitations substantially modifies the density and spin response
of the system. Some of the modifications are not very obvious. At the end of the Sec.~\ref{sec:kp} we
already mentioned the significant difference between spin and density responses in the long wavelength limit.
Let us consider this in more detail. The spin response of the noninteracting electron gas coincides with the
density response and can be expressed through the Lindhard function. The spin-orbit interaction within the
valence band breaks down this correspondence.

In Fig.~\ref{fig2} we plot the imaginary part of the noninteracting
density and longitudinal spin response functions in $p$-doped GaAs for different wave vectors. For a small
wave vector $q=0.003\mbox{\AA}^{-1}$ the longitudinal spin response exhibits a strong peak around
0.2 eV associated with inter-valence band spin excitations between heavy and light hole subbands. The
corresponding density excitations are suppressed due to the orthogonality of the initial and final states,
see Eq.~(\ref{imchi}). As a result, the density response for short wavevectors is almost nonexistent.
If we increase the wave vector to $q=0.05{\mbox \AA}^{-1}$, the intraband excitations within the heavy hole band
become noticeable in both density and spin responses. The longitudinal spin response, however, still
prevails in the range of inter-valence band transitions.

This leads us to conclude that the long-wavelength spectrum of the single-particle excitations in $p$-doped
semiconductors is dominated by the inter-valence band spin excitations. The origin of this effect is in the
spin-orbit interaction, which mixes spin and orbital degrees of freedom. Without the spin-orbit interaction,
vertical spin excitations would be prohibited due to the orthogonality of the orbital parts of Bloch functions.

\begin{figure}
\centering
\includegraphics[width=0.95\linewidth]{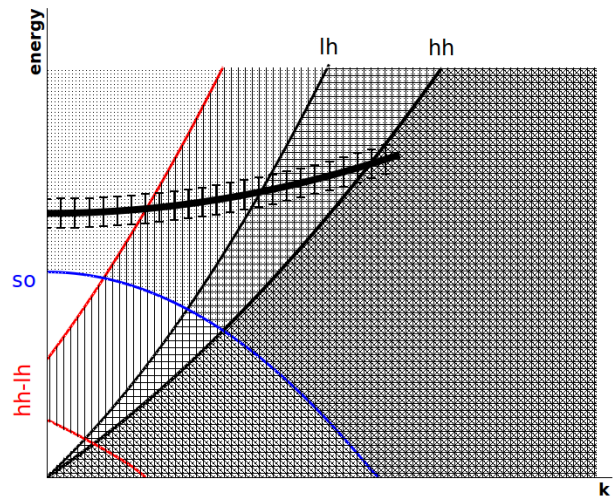}
\caption{(Color online) Schematic diagram of the excitation spectrum within the semiconductor valence band.
Labels indicate the edges of single-particle excitation regions within the heavy hole band (hh), within the
light hole band (lh), between heavy hole and light hole bands (hh-lh) and between the split-off band and
heavy and light hole bands (so), see Fig.~\ref{fig1}. As a result, the plasmon mode in the valence
band lies entirely within the single-particle excitation spectrum and is effectively suppressed due to
Landau damping. } \label{fig3}
\end{figure}

Another interesting feature of $p$-doped semiconductors is the effective suppression of the collective modes
in the valence band. In the conventional picture of the conduction band, collective plasmon excitations
are well defined in the long-wavelength side of the excitation spectrum. With increasing momentum,
the collective mode approaches and then enters the region of single-particle excitations, where it becomes
rapidly suppressed due to Landau damping.

\begin{figure}
\centering
\includegraphics[width=0.95\linewidth]{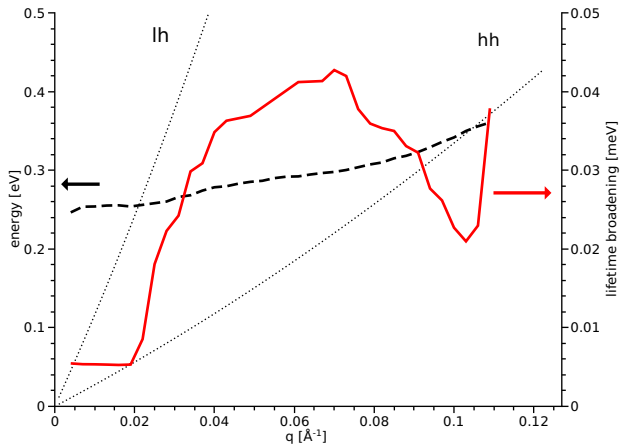}
\caption{(Color online) Dispersion (dashed black) and lifetime broadening (solid red) of the valence
band plasmon calculated for the $p$-doped GaAs with the hole concentration of $p=3.5\times 10^{20} \; {\rm cm}^{-3}$.
Dotted lines correspond to the onset of the intraband single-particle excitations within the light hole and heavy
hole bands.} \label{fig4}
\end{figure}

The situation is different for the valence band. In Fig.~\ref{fig3} we plot a schematic diagram of
the excitation spectrum. The excitation region for single-particle transitions within the heavy hole band is
qualitatively similar to that of the conduction band. In the valence band, however, the single-particle
excitation spectrum is extended due to the intraband transitions within the light hole band and interband
transitions between heavy and light hole bands and between split-off and heavy/light hole bands
(red and blue arrows in Fig.~\ref{fig1}). In Fig.~\ref{fig3} the corresponding regions of single-particle
excitations are shaded with different patterns. It can be seen that the collective mode in the valence
band falls entirely within the region of single-particle excitations and, therefore, becomes suppressed even at the
long-wavelength side of the spectrum. Error bars in Fig.~\ref{fig3} indicate the plasmon resonance broadening due to Landau damping.

\begin{figure}
\centering
\includegraphics[width=0.95\linewidth]{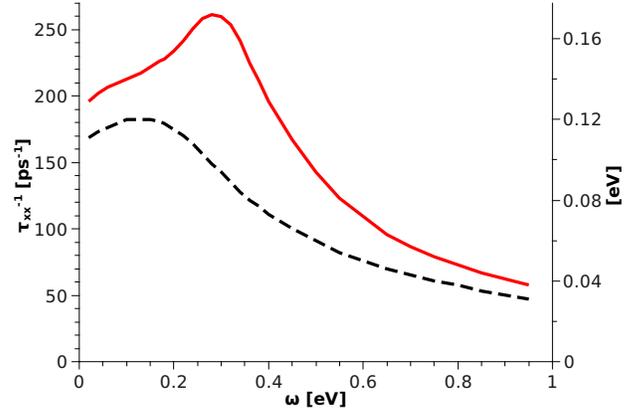}
\caption{(Color online) Total (charge and spin) carrier relaxation rate for
Ga$_{\rm 0.948}$Mn$_{\rm 0.052}$As with hole concentration $p=3\times 10^{20} \; {\rm cm}^{-3}$.
Dashed line: static screening model. Solid line:
evaluation of Eq.~(\ref{tau}) with full dynamic TDDFT treatment of electron interaction.
See discussion in text.} \label{fig5}
\end{figure}

To illustrate the effect we have performed numerical calculations of the plasmon dispersion and the
lifetime broadening of the collective excitations in the valence band of the $p$-doped GaAs. The plasmon
frequencies were determined as the zeros of the real part of the RPA dielectric function and the lifetime
broadening is associated with the imaginary part of the frequency poles. In Fig.~\ref{fig4} the black and
red lines correspond to the dispersion and the lifetime of the plasmon excitations, respectively.
The dotted lines indicate the regions of the intraband single-particle excitations within the light hole
and heavy hole bands, compare with Fig.~\ref{fig3}. At small wavevectors the plasmon mode falls within
the region of inter-valence band single-particle excitations resulting in a lifetime broadening of the
collective resonance of about 5 meV. Once the plasmon dispersion enters the region of single-particle
excitations within the light hole band, the life-time broadening substantially increases into the 30-40 meV range.
An additional sharp rise in the damping takes place when the collective mode enters the region of heavy hole
intraband excitations.

\subsection{Magnetically doped GaMnAs}\label{sec:doped}

The introduction of magnetic impurities in GaAs has two consequences. First, charge and spin
disorder are brough into the system and, second, the mean field part of the $p$-$d$ exchange interaction between localized
spins and itinerant holes causes modifications of the valence band structure once the system
enters the magnetically ordered phase.

\begin{figure}
\centering
\includegraphics[width=0.95\linewidth]{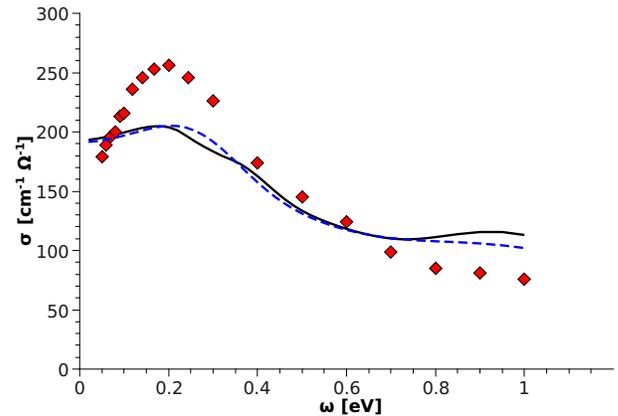}
\caption{(Color online) Infrared conductivity of ferromagnetic Ga$_{\rm 0.948}$Mn$_{\rm 0.052}$As with
hole concentration $p=3\times 10^{20} \; {\rm cm}^{-3}$. Calculations are performed according to Eq.~(\ref{self}),
and using a relaxation rate obtained through Eq.~(\ref{tau}) (solid line) or a fixed $\tau^{-1}=230\,
{\rm ps}^{-1}$ (dashed line). Symbols are the experimental data of Ref.~\onlinecite{singley}.} \label{fig6}
\end{figure}

Let us consider the effect of disorder first. In calculating carrier relaxation rates, most theoretical
models for GaMnAs use a static screening approach, where all many-body effects are reduced to the static
screening of the Coulomb disorder potential. Within our model, however, the momentum and frequency dependent
relaxation rate (\ref{tau}) is expressed through the set of density and spin-density response functions that
allows us to use the full dynamic treatment of electron-electron interaction, thus accounting for the variety of
many-body effects including correlations and collective modes.

In Fig.~\ref{fig5} we plot the frequency dependence
of the total (charge and spin) relaxation rate calculated for Ga$_{\rm 0.948}$Mn$_{\rm 0.052}$As within the
static screening model and using the full dynamic treatment of electron-electron interaction according to Eq.~(\ref{tau}).
The difference between the two curves in the static limit is due to the xc part of the electron-electron
interaction that affects both charge and spin scattering. The most striking difference, however, is the pronounced
feature appearing between 0.2 eV and 0.5 eV associated with the collective modes. Although we have seen above
that the collective excitations are significantly damped in the valence band, they still play an important role
in the transport properties of the system providing an effective channel for momentum relaxation. Their
contributions give up to 50\% increase to the total carrier relaxation rate. Note that, due to their longitudinal character,
the plasmon modes do not directly affect the optical response and enter only
indirectly through the tensor of frequency and momentum dependent relaxation rates (\ref{tau}).

\begin{figure}
\centering
\includegraphics[width=0.95\linewidth]{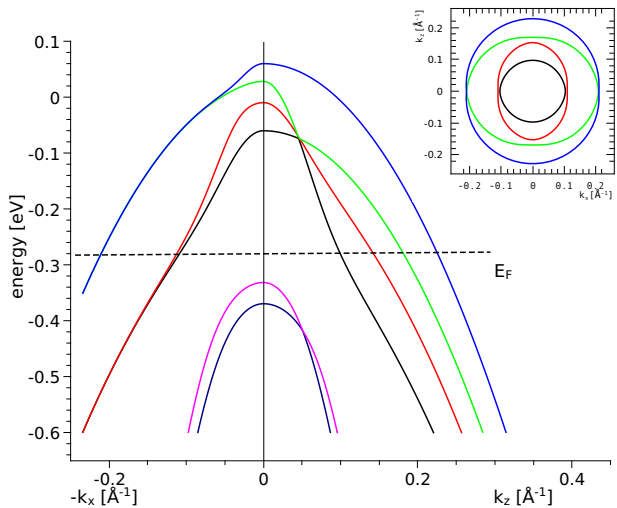}
\caption{(Color online) Band structure of ferromagnetic Ga$_{\rm 0.95}$Mn$_{\rm 0.05}$As with hole
concentration $p=3.5\times 10^{20} \; {\rm cm}^{-3}$. Alignment of localized spins results in strongly anisotropic
valence band spin splitting. Inset shows a cut of the Fermi surface by the plane $k_y=0$. } \label{fig7}
\end{figure}

In Fig.~\ref{fig6} we compare our calculations of the infrared conductivity of ferromagnetic
Ga$_{\rm 0.948}$Mn$_{\rm 0.052}$As with the experimental data of Singley {\em et al.} \cite{singley} The calculations
were performed according to Eq.~(\ref{self}). Solid line corresponds to a relaxation rate obtained through
Eq.~(\ref{tau}), dashed line describes calculations with the fixed $\tau^{-1}=230\, {\rm ps}^{-1}$. The theory
shows qualitative agreement with the experiment. The insensitivity of the calculations to the frequency
dependence of relaxation rate (minor difference between solid and dashed lines in Fig.~\ref{fig6}) suggests
that effects of the band structure play the dominant role in determining the shape of the infrared conductivity
and overshadow the strong frequency dependence of $\tau$ obtained within our model and presented in Fig.~\ref{fig5}.

An alternative possible experimental probe that could reveal the details of the frequency and momentum dependence of the
carrier relaxation rate in more explicit ways are measurements of the position and lineshape of the
plasmon resonance itself. It was shown in Ref.~\onlinecite{belitz} that these quantities are sensitive to the
carrier relaxation time, with both real and imaginary part of $\tau$ and its dynamic nature being essential.
Our approach seems to fit well to describe such experiments.

\begin{figure}
\centering
\includegraphics[width=0.95\linewidth]{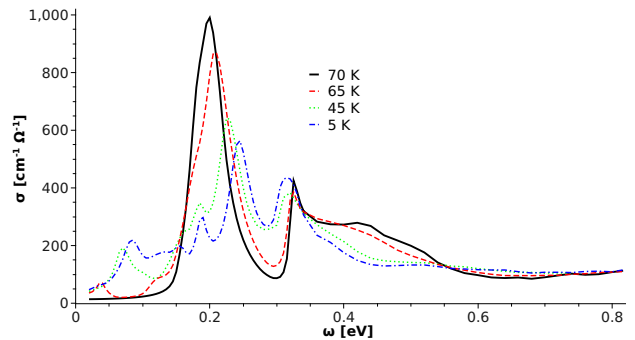}
\caption{(Color online) Temperature dependence of infrared conductivity of Ga$_{\rm 0.948}$Mn$_{\rm 0.052}$As with
hole concentration $p=3\times 10^{20} \; {\rm cm}^{-3}$ and $T_c=70$ K calculated with weak disorder,
lifetime broadening of $\Gamma=5$ meV.} \label{fig7a}
\end{figure}

As was mentioned before, the magnetic impurities bring localized spins into the system, which interact with
the itinerant carriers through the $p$-$d$ exchange interaction. The fluctuating part of this interaction
constitutes the spin disorder. The mean field part of exchange interaction, which we absorb into the clean system
band structure Hamiltonian $\hat{H}_e$, is responsible for the spin splitting of the valence bands once the system
enters the magnetically ordered state. Due to the spin-orbit interaction within the valence band, this spin splitting
strongly depends both on the magnitude and direction of the wave vector ${\bf k}$.

In Fig.~\ref{fig7} we plot the
band structure of ferromagnetic Ga$_{\rm 0.95}$Mn$_{\rm 0.05}$As. Strong anisotropy of the valence band
spin-splitting is seen between directions along and perpendicular to the magnetization of localized spins ($z$-direction).
The inset shows a cut of the Fermi surface by the plane $k_y=0$. One can easily see the distortion of the Fermi surface
from the spherical shape of the paramagnetic system (for clarity we have neglected here the valence
band warping, but it is included in our calculations). The modification of the Fermi surface together with the suppression
of localized spin fluctuations
are responsible for the significant drop in static resistivity of GaMnAs during the transition from paramagnetic
to ferromagnetic state. This effect was considered before.\cite{lopez,KUPRB09}

\begin{figure}
\centering
\includegraphics[width=0.95\linewidth]{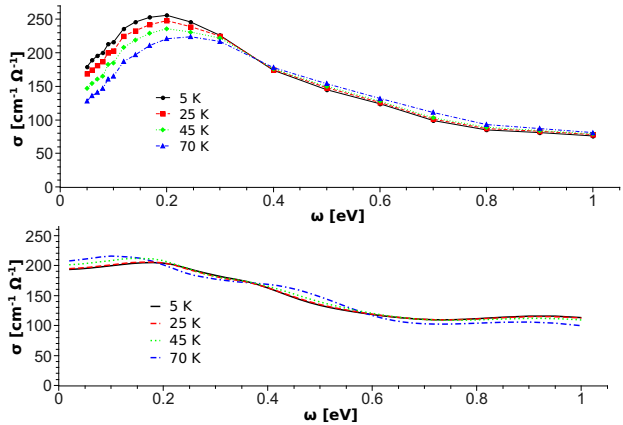}
\caption{(Color online) Temperature dependence of the infrared conductivity of Ga$_{\rm 0.948}$Mn$_{\rm 0.052}$As with
hole concentration $p=3\times 10^{20} \; {\rm cm}^{-3}$ and $T_c=70$ K. Upper panel: experimental data of
Ref.~\onlinecite{singley}. Lower panel: results from Eq.~(\ref{self}).} \label{fig8}
\end{figure}

Here we point out that the modification of the valence band structure during the transition from
paramagnetic to ferromagnetic state also modifies energies and oscillator strengths of intervalence band optical
transitions affecting thus the infrared conductivity as well. To better show the underlying physics of temperature
induced changes, we plot in Fig.~\ref{fig7a} the infrared conductivity for the sample parameters of Ref.~\onlinecite{singley},
but with a small lifetime broadening of $\Gamma=5$ meV. In the paramagnetic state (solid line) three features
can be identified: a strong peak around 0.2 eV corresponding to the heavy hole - light hole transitions, a smaller
peak with a broad shoulder around 0.4 eV associated with the split off to light hole transitions and a wide
background of split off to heavy hole transitions.

With the temperature going below $T_c=70$ K, two main
phenomena occur. The first is the suppression of the high energy shoulder of the split off to light hole
transitions. The second is the appearance of the transitions between the spin-split heavy hole and light
hole bands and the redistribution of the oscillator strength among them. The lowest energy peaks correspond to
the transitions between spin-split bands. Calculations were performed for light linearly polarized
in the plane perpendicular to the magnetization. Due to the spin-orbit interaction within the valence band,
the transitions between the spin-split states are optically allowed. The additional peak at higher energy
corresponds to heavy hole-light hole ``spin-flip'' transitions. As the temperature goes down, the spin
splitting increases and the ``spin-flip'' transitions gain the intensities at the account of ``spin-conserving''
heavy hole-light hole transitions.

The real GaMnAs samples are much more disordered. In Fig.~\ref{fig8} we compare experimental data on infrared
conductivity of Ga$_{\rm 0.948}$Mn$_{\rm 0.052}$As from Ref.~\onlinecite{singley} with calculations
using our model of Eqs.~(\ref{self}) and (\ref{tau}). The large disorder induced life-time broadening blankets
most of the features discussed above. The suppression of the high energy shoulder of split off to light hole
transitions in the ferromagnetic state is seen, however, both on the experimental and theoretical plots. Overall,
for energies above the main peak position around 0.2 eV, the calculations are in good agreement with the
experimental results.

Note also that unlike in Ref.~\onlinecite{ewelina}, our calculations do not require
incorporation of an impurity band within the energy gap to avoid a drop in conductivity around 0.8-1 eV.
At energies below the main peak position the agreement with the experiment is worse. We should mention, however,
that this is the region of $\omega \tau \leq 1$ where our calculations are less reliable due to approximate
nature of the expression (\ref{self}). The self-consistent evaluation of Eq.~(\ref{finalS}) should be used
there instead. Once the frequency goes to zero, the static conductivity should more appropriately be calculated using an expression derived
from the semiclassical Boltzmann equation.\cite{rates} We have investigated this regime before\cite{KUPRB09}
to describe the drop in static resistivity in the ferromagnetic phase.

\section{Conclusions}\label{sec:conclusion}
We have developed a comprehensive theory of transport in spin and charge disordered media. The theory is based on the
equation of motion of the paramagnetic current response function of the disordered system, treats disorder
and many-body effects on equal footings, and combines a ${\bf k\cdot p}$ based description of the semiconductor
valence band structure with a full dynamic treatment of electron-electron interaction by means of TDDFT. We have applied our
theory to the specific case of GaMnAs.

We have shown that the multiband nature and spin-orbit interaction within the valence band bring new effects
for p-doped GaAs as compared to the conventional n-type systems. The density and spin-density responses of
noninteracting carriers within the valence band are not the same anymore. Moreover, the long wavelength side
of the single-particle excitation spectrum is now completely dominated by the inter-valence band {\it spin}
excitations. Due to the extended region of single-particle excitations within the valence band, the collective
plasmon mode entirely falls within the region of these excitations and, therefore, is effectively damped for all wave vectors.

For the magnetically doped system the mean-field part of the $p$-$d$ exchange interaction between itinerant
holes and localized spins substantially modifies the semiconductor band structure once the system enters a
magnetically ordered phase. This modification substantially affects energies and oscillator strengths of
the intervalence band optical transitions. Our calculations are in good agreement with
experimental data for the temperature dependence of the infrared conductivity in GaMnAs.

A full dynamical treatment of electron-electron interactions is essential to capture the influence of the
collective excitations on the carrier relaxation rate. Our calculations show that, by providing an effective
channel of momentum relaxation, the collective excitations within the valence band significantly (up to 50\%)
increase the transport relaxation rate.

However, it turns out that the actual infrared absorption spectra are not very sensitive to the details of the frequency
dependence of the relaxation rate, but are mostly determined by the features of the band structure. Direct
measurements of the position and lineshape of the plasmon resonance itself are likely to be more sensitive to
the details of the frequency and momentum dependence of the carrier relaxation rate.

The theory presented here, treating disorder and many-body effects on
equal footings, provides a very general framework for describing electron dynamics in materials. It can,
in principle, be made self-consistent and thus be applied beyond the weak-disorder limit;
it can accommodate many different types of disorder, as well as band structure models.
This should make it well suited for further exploration of the optical and transport properties of DMSs and
other systems of practical interest.

\begin{acknowledgments}

This work was supported by DOE under Grant No. DE-FG02-05ER46213.

\end{acknowledgments}

\appendix

\section{Generalized ${\bf k\cdot p}$ approach}\label{app:kp}
The derivation of the generalized ${\bf k\cdot p}$ perturbation approach presented here is based on Ref. \onlinecite{birpikus}.
First, the electronic wave function is expanded in the Luttinger-Kohn basis\cite{LK}
\begin{equation}\label{expansion}
  \Psi=\sum_{n,{\bf k}}A_n({\bf k})\chi_{n,{\bf k}}= \frac1{\sqrt{V}}\sum_{n,{\bf k}}A_n({\bf k})e^{i{\bf k r}} |u_{n,0}\rangle,
\end{equation}
where $|u_{n,0}\rangle$ are periodic parts of Bloch functions at ${\bf k}=0$, and $A_n({\bf k})$ are the expansion coefficients.
This results in the following matrix form of the Schr\"odinger equation:
\begin{equation}\label{kp}
  \sum_{n,{\bf k}} A_n({\bf k})\left[ \left(\varepsilon_{n,0}+\frac{\hbar^2 k^2}
  {2 m_0}-\varepsilon\right)\delta_{n',n}
  +\frac{\hbar}{m_0}{\bf k\cdot \pi}_{n',n}\right]=0,
\end{equation}
where $\varepsilon_{n,0}$ are the band edge energies at ${\bf k}=0$.

The last term in Eq.~(\ref{kp}) mixes states with different $n$ for ${\bf k}\neq 0$. Now we separate the
whole set of the bands $\{n\}$ into those whose contribution we are going to calculate exactly $\{s\}$, and
the remote bands $\{r\}$ that we will treat up to the second order in momentum. Equation (\ref{kp}) can be represented as
\begin{equation} \label{SEcan}
(H_0+H_1+H_2){\bf A}=\varepsilon {\bf A},
\end{equation}
where ${\bf A}$ is the vector of coefficients $A_n({\bf k})$, $H_0$ is the diagonal part of Hamiltonian,
and $H_1$ and $H_2$ correspond to the block-diagonal and off-block-diagonal parts of the ${\bf k\cdot \pi}$ matrix
with respect to the included and remote bands.
Next, we apply the canonical transformation
\begin{equation}\label{canon}
{\bf A}= e^{S} {\bf B}=e^{S_1+S_2} {\bf B},
\end{equation}
with $S_1$ and $S_2$ being antihermitian operators of first and second order in the perturbation, respectively. The matrix equation
(\ref{SEcan}) then has the form
\begin{equation}
\left\{ e^{-S_1-S_2}(H_0+H_1+H_2)e^{S_1+S_2}\right\}{\bf B}=
\bar{H}{\bf B}=\varepsilon {\bf B}.
\end{equation}
By choosing
\begin{equation}
H_2 + [H_0,S_1]=0, \;\;\;\;\; [H_0,S_2]+[H_1,S_1]=0,
\end{equation}
where $[...]$ denotes the commutator, we write up to terms of second order in the perturbations $H_1$ and $H_2$
\begin{equation}\label{Hbar}
 \bar{H}\approx H_0+H_1+\frac12 [H_2,S_1].
\end{equation}
The matrix elements between the Luttinger-Kohn periodic amplitudes $|u_{n,0}\rangle \equiv |n\rangle$ are
\begin{eqnarray}\label{kpfirst}
\ds \langle n| H_0 | n'\rangle & = & \left( \varepsilon_{n,0}+\frac{\hbar^2 k^2}{2 m_0}\right)\delta_{n,n'}, \\
\ds \langle s| H_1 | s'\rangle & = & \sum_{\alpha}\frac{\hbar k_{\alpha}\pi^{\alpha}_{s,s'}}{m_0},\\
\ds \langle s| H_2 | r\rangle & = & \sum_{\alpha}\frac{\hbar k_{\alpha}\pi^{\alpha}_{s,r}}{m_0}, \\
\ds \langle s| S_1 | r\rangle & = & -\frac{\langle s| H_2 | r\rangle} {\langle s| H_0 | s\rangle -
 \langle r| H_0 | r\rangle}
 \nonumber\\
 &&
 = \sum_{\alpha}\frac{\hbar k_{\alpha}\pi^{\alpha}_{s,r}}{m_0}\frac{1}
 {\varepsilon_{r,0}-\varepsilon_{s,0}}.  \label{S1}
\end{eqnarray}
For the last term in (\ref{Hbar}) we can then write
\begin{eqnarray}
\langle s| [H_2,S_1] | s'\rangle  &=&  \sum_{r} \Big\{ \langle s| H_2 | r\rangle \langle r| S_1 | s' \rangle
\nonumber\\
&& {} - \langle s| S_1 | r \rangle \langle r| H_2 | s'\rangle \Big\}
 \label{kplast}  \\
& = & \sum_{\alpha,\beta \atop r} \frac{\hbar^2 k_{\alpha}k_{\beta}}{m_0^2} \!
\left(\frac{\pi^{\alpha}_{s,r}\pi^{\beta}_{r,s'}}{\varepsilon_{s',0} -\varepsilon_{r,0}}+\frac
{\pi^{\beta}_{s,r}\pi^{\alpha}_{r,s'}}{\varepsilon_{s,0}-\varepsilon_{r,0}}\right) \! .  \nonumber \label{rem_bands}
\end{eqnarray}
Here we used the fact that the $H_2$ and $S_1$ operators have only off-block-diagonal matrix elements between
the $s$ and $r$ bands. Eqs.~(\ref{kpfirst})-(\ref{kplast}) define the matrix of the effective Hamiltonian (\ref{Hbar}).
Nonvanishing matrix elements are determined by the symmetry of the crystal.

\begin{widetext}

\section{Evaluation of the matrix elements in Eqs.~(\ref{currentMB}) and (\ref{spindensityMB})}\label{app:me}
In order to evaluate Eq.~(\ref{currentMB}) we need to calculate the following matrix element:
\begin{equation}\label{melement}
\hbar\left(k_{\alpha}-\frac{q_{\alpha}}2\right) \langle u_{i',{\bf k-q}}  | u_{i,{\bf k}}\rangle+
  \langle u_{i',{\bf k-q}} |\hat{\pi}_{\alpha} | u_{i,{\bf k}}\rangle=
\left\langle u_{i',{\bf k-q}} \left| \hbar\left(k_{\alpha}-\frac{q_{\alpha}}2\right)+ \hat{\pi}_{\alpha} \right| u_{i,{\bf
k}}\right\rangle,
\end{equation}
where $|u_{i,{\bf k}}\rangle$ is expressed through the amplitudes at the zone center:
\begin{equation}\label{uik}
|u_{i,{\bf k}}\rangle=\sum_{n}A_n(i,{\bf k}) |u_{n,0}\rangle.
\end{equation}

From diagonalization of the effective Hamiltonian (\ref{Hbar}), however, we obtain coefficients $B_n(i,{\bf k})$ related
to $A_n(i,{\bf k})$ through Eq.~(\ref{canon}). Expanding $e^{S}\approx 1+S$, we express
\begin{equation}
  |u_{i,{\bf k}}\rangle=\sum_{s}B_s(i,{\bf k}) |s\rangle+\sum_s \sum_r \langle r|S({\bf k})|s\rangle B_s(i,{\bf k}) |r\rangle,
\end{equation}
where we have used the fact that the coefficients $B_n$ are non-zero only for exact bands and $S$ has only off-block-diagonal matrix elements.
The bra vector is
\begin{equation}
  \langle u_{i',{\bf k'}}|=\sum_{s'}B^*_{s'}(i',{\bf k'}) \langle s' |-
  \sum_{s'} \sum_{r'} \langle s'|S({\bf k'})|r'\rangle B^*_{s'}(i',{\bf k'}) \langle r' |,
\end{equation}
where we have used the antihermiticity of $S$. Matrix elements of an arbitrary operator $\hat{F}$ to the lowest order in $S$ can then
be expressed as follows:
\begin{equation}
  \langle u_{i',{\bf k'}}|\hat{F}|u_{i,{\bf k}}\rangle =\sum_{s' s} B^*_{s'}(i',{\bf k'}) B_s(i,{\bf k})
  \left( \langle s' |\hat{F}|s\rangle+\sum_{r}\left(\langle s' |\hat{F}|r\rangle \langle r|S({\bf k})|s\rangle -
  \langle s' |S({\bf k'})|r\rangle \langle r|\hat{F}|s\rangle \right)\right).
\end{equation}
Using Eq.~(\ref{S1}) for matrix elements of $\hat{S}_1$, we have
\begin{equation}\label{matF}
  \langle u_{i',{\bf k'}}|\hat{F}|u_{i,{\bf k}}\rangle =\sum_{s' s} B^*_{s'}(i',{\bf k'}) B_s(i,{\bf k})
  \left( \langle s' |\hat{F}|s\rangle-\frac{\hbar}{m_0}\sum_{\lambda,r}\left(
  \frac{k_{\lambda}\langle s' |\hat{F}|r\rangle \langle r|\hat{\pi}^{\lambda}|s\rangle}{\varepsilon_r-\varepsilon_s}+
  \frac{k'_{\lambda}\langle s'|\hat{\pi}^{\lambda}|r\rangle \langle r |\hat{F}|s\rangle }{\varepsilon_r-\varepsilon_{s'}}
  \right)\right).
\end{equation}
The matrix element (\ref{melement}) has thus the following form:
\begin{eqnarray*}
  \left\langle u_{i',{\bf k-q}} \left| \hbar\left(k_{\alpha}-\frac{q_{\alpha}}2\right)+ \hat{\pi}_{\alpha} \right| u_{i,{\bf
k}}\right\rangle &=& \sum_{s' s} B^*_{s'}(i',{\bf k-q}) B_s(i,{\bf k}) \\
  &\times& \left[ \hbar\left(k_{\alpha}-\frac{q_{\alpha}}2\right)\delta_{s's}+\langle s' |\hat{\pi}^{\alpha}|s\rangle+
  \frac{\hbar}{m_0}\sum_{\lambda,r}\left(
  \frac{k_{\lambda}\pi^{\alpha}_{s',r}\pi^{\lambda}_{r,s}}{\varepsilon_s-\varepsilon_r}+
  \frac{(k_{\lambda}-q_{\lambda})\pi^{\lambda}_{s',r}\pi^{\alpha}_{r,s} }{\varepsilon_{s'}-\varepsilon_r}
  \right)\right].
\end{eqnarray*}
For ${\bf q}=0$ it reduces to
\begin{equation}
  \left\langle u_{i',{\bf k}} \left| \hbar k_{\alpha} + \hat{\pi}_{\alpha} \right| u_{i,{\bf
k}}\right\rangle = \sum_{s' s} B^*_{s'}(i',{\bf k}) B_s(i,{\bf k})
  \left[ \hbar k_{\alpha}\delta_{s's}+\langle s' |\hat{\pi}^{\alpha}|s\rangle+
  \frac{\hbar}{m_0}\sum_{\lambda,r}k_{\lambda}\left(
  \frac{\pi^{\alpha}_{s',r}\pi^{\lambda}_{r,s}}{\varepsilon_s-\varepsilon_r}+
  \frac{\pi^{\lambda}_{s',r}\pi^{\alpha}_{r,s} }{\varepsilon_{s'}-\varepsilon_r}
  \right)\right].
\end{equation}
By comparison with the expressions derived in Appendix A, we find that this reduces to
\begin{equation}
  \left\langle u_{i',{\bf k}} \left| \hbar k_{\alpha} + \hat{\pi}_{\alpha} \right| u_{i,{\bf
k}}\right\rangle = \sum_{s' s} B^*_{s'}(i',{\bf k}) B_s(i,{\bf k})
\frac{m_0}{\hbar} \frac{\de}{\de k_{\alpha}} \langle s' |\bar{H}| s \rangle,
\end{equation}
where $\bar{H}$ is the Hamiltonian (\ref{Hbar}).

The matrix elements of the spin operator in Eq.~(\ref{spindensityMB}) should also be evaluated through Eq.~(\ref{matF}):
\begin{equation}
  \langle u_{i',{\bf k'}}|\hat{\sigma}^{\mu}|u_{i,{\bf k}}\rangle =\sum_{s' s} B^*_{s'}(i',{\bf k'}) B_s(i,{\bf k})
  \left( \langle s' |\hat{\sigma}^{\mu}|s\rangle-\frac{\hbar}{m_0}\sum_{\lambda,r}\left(
  \frac{k_{\lambda}\langle s' |\hat{\sigma}^{\mu}|r\rangle \langle r|\hat{\pi}^{\lambda}|s\rangle}{\varepsilon_r-\varepsilon_s}+
  \frac{k'_{\lambda}\langle s'|\hat{\pi}^{\lambda}|r\rangle \langle r |\hat{\sigma}^{\mu}|s\rangle }{\varepsilon_r-\varepsilon_{s'}}
  \right)\right).
\end{equation}

\end{widetext}

Let us look now at the sum over remote bands. Since the spin operator acts only on the spin part of the basis functions, only
those remote bands whose orbital part has the same symmetry as the exact bands will contribute to this sum.

If we are considering a $6\times 6$
Hamiltonian and neglect inversion asymmetry, the exact states are $p$-bonding states that transform according to the $F_1^+$
representation of the point group $O_h$ ($\Gamma_{15}^{\prime}$ small representation). The momentum operator transforms as $F_2^-$,
and since the direct product $F_1^+ \times F_2^- \times F_1^+$ does not contain a unit representation, the sum over remote bands
vanishes. There may be a small contribution in $T_d$ crystals, but it can be considered negligible.

If we are working in an
8-band ${\bf k\cdot p}$ model, there are possible contributions to the sum when $|s\rangle$ and $|r\rangle$ are
$\Gamma_1^{\prime}$ states and $|s'\rangle$ is $\Gamma_{15}^{\prime}$ and vice versa. Since there is only a small
admixture of the conduction band amplitude to the
valence band states, these contributions are expected to be small and therefore can be neglected.

Based on this reasoning, we use the following approximation:
\begin{equation}
  \langle u_{i',{\bf k'}}|\hat{\sigma}^{\mu}|u_{i,{\bf k}}\rangle \approx \sum_{s' s} B^*_{s'}(i',{\bf k'}) B_s(i,{\bf k})
  \langle s' |\hat{\sigma}^{\mu}|s\rangle.
\end{equation}

\section{$8\times 8$ Hamiltonian} \label{app:8band}
In the basis
\begin{eqnarray}
  |1\rangle &=& |E,+\frac12\rangle=S\up, \nonumber \\
  |2\rangle &=& |E,-\frac12\rangle=iS\dn, \nonumber\\
  |3\rangle &=& |HH,+\frac32\rangle=\frac1{\sqrt2}(X+iY)\up, \nonumber\\
  |4\rangle &=& |LH,+\frac12\rangle=\frac{i}{\sqrt6}[(X+iY)\dn-2Z\up], \nonumber\\
  |5\rangle &=& |LH,-\frac12\rangle=\frac1{\sqrt6}[(X-iY)\up+2Z\dn], \label{basisKane}\\
  |6\rangle &=& |HH,-\frac32\rangle=\frac{i}{\sqrt2}(X-iY)\dn, \nonumber\\
  |7\rangle &=& |SO,+\frac12\rangle=\frac1{\sqrt3}[(X+iY)\dn+Z\up], \nonumber\\
  |8\rangle &=& |SO,-\frac12\rangle=\frac{i}{\sqrt3}[-(X-iY)\up+Z\dn], \nonumber
\end{eqnarray}
the Hamiltonian matrix has the form
\begin{widetext}
\begin{equation}\label{H8x8}
  \left(   \begin{array}{cccccccc}
  \ds E_g+\frac{\hbar^2 k^2}{2\widetilde{m}_e} & 0 & \ds \frac{i}{\sqrt2}V k_+ & \ds \sqrt{\frac23}V k_z &
  \ds \frac{i}{\sqrt6}V k_- & 0 & \ds \frac{i}{\sqrt3}V k_z & \ds \frac{1}{\sqrt3}V k_- \vspace{0.2cm}\\
  0 & \ds E_g+\frac{\hbar^2 k^2}{2\widetilde{m}_e} & 0 & \ds \frac{i}{\sqrt6}V k_+ & \ds \sqrt{\frac23}V k_z &
  \ds \frac{i}{\sqrt2}V k_- & \ds \frac{1}{\sqrt3}V k_+ & \ds \frac{i}{\sqrt3}V k_z \vspace{0.2cm}\\
  \ds -\frac{i}{\sqrt2}V k_- & 0 & P+Q & L & M & 0 & \ds \frac{i}{\sqrt2}L' & -i\sqrt2 M' \vspace{0.2cm}\\
  \ds \sqrt{\frac23}V k_z & \ds -\frac{i}{\sqrt6}V k_- & L^* & P-Q & 0 & M & -i\sqrt2 Q' &
   \ds i \sqrt{\frac32}L' \vspace{0.2cm}\\
  \ds -\frac{i}{\sqrt6}V k_+ & \ds \sqrt{\frac23}V k_z & M^* & 0 & P-Q & -L &  \ds -i \sqrt{\frac32}L^{\prime *} &
   -i\sqrt2 Q' \vspace{0.2cm}\\
  0 & \ds -\frac{i}{\sqrt2}V k_+ & 0 & M^* & -L^* & P+Q & -i\sqrt2 M'^* & \ds -\frac{i}{\sqrt2} L'^{*} \vspace{0.2cm}\\
  \ds -\frac{i}{\sqrt3}V k_z & \ds \frac{1}{\sqrt3}V k_- & \ds -\frac{i}{\sqrt2}L'^* & i\sqrt2 Q' &
  \ds i \sqrt{\frac32}L' & i\sqrt2 M' & P'-\Delta & 0 \vspace{0.2cm}\\
  \ds \frac{1}{\sqrt3}V k_+ & \ds -\frac{i}{\sqrt3}V k_z & i\sqrt2 M'^* & \ds -i \sqrt{\frac32}L'^* & i\sqrt2 Q' &
  \ds \frac{i}{\sqrt2} L' & 0 & P'-\Delta
  \end{array} \right)
\end{equation}
\end{widetext}
with
\begin{eqnarray*}
k_{\pm} &=& k_x\pm i k_y, \\
V &=& -i\frac{\hbar}{m_0}\langle S|\hat{p}_x|X\rangle=\sqrt{E_p\frac{\hbar^2}{2 m_0}}.
\end{eqnarray*}
Interaction with remote bands results in the intra valence band terms
\begin{eqnarray*}
P^{(\prime)} &=& -\frac{\hbar^2}{2m_0} \widetilde{\gamma}^{(\prime)}_1 k^2, \\
Q^{(\prime)} &=& -\frac{\hbar^2}{2m_0} \widetilde{\gamma}^{(\prime)}_2 (k_x^2+k_y^2-2k_z^2), \\
L^{(\prime)} &=& \frac{\hbar^2}{2m_0} i2\sqrt3\widetilde{\gamma}^{(\prime)}_3 k_z k_-, \\
M^{(\prime)} &=& -\frac{\hbar^2}{2m_0} \sqrt3 [\widetilde{\gamma}^{(\prime)}_2 (k_x^2-k_y^2)- i\widetilde{\gamma}^{(\prime)}_3 (k_x k_y+k_y k_x) ],
\end{eqnarray*}
where renormalization leads to
\begin{eqnarray*}
\frac1{\widetilde{m}_e} & = & \frac1{m^*_e}-\frac1{m_0}\frac{E_p}3
\left(\frac2{E_g}+\frac1{E_g+\Delta}\right), \\
\widetilde{\gamma}_1 & = & \gamma_1-\frac{E_p}{3E_g}, \\
\widetilde{\gamma}'_1 &=& \gamma_1-\frac{E_p}{3(E_g+\Delta)}, \\
\widetilde{\gamma}_2 & = & \gamma_2-\frac{E_p}{6E_g}, \\
\widetilde{\gamma}'_2 &=&
\gamma_2-\frac{E_p}{12}\left(\frac1{E_g}+\frac1{E_g+\Delta}\right),\\
\widetilde{\gamma}_3 & = & \gamma_3-\frac{E_p}{6E_g}, \\
\widetilde{\gamma}'_3 &=&
\gamma_3-\frac{E_p}{12}\left(\frac1{E_g}+\frac1{E_g+\Delta}\right).
\end{eqnarray*}
This reflects the fact that the interaction between conduction and valence bands is taken in our Hamiltonian explicitly.
In writing the matrix (\ref{H8x8}) we have neglected small terms associated with the lack of inversion symmetry in $T_d$ crystals.

The matrix of the mean-field part of the $s(p)$-$d$ exchange interaction, which is responsible for the band spin splitting in
the magnetically ordered phase, has the form
\begin{widetext}
\begin{equation}
-\frac12 \langle S \rangle x N_0 \left( \begin{array}{cccccccc}
 \hspace{3mm} \alpha\hspace{3mm} & \hspace{3mm}0\hspace{3mm} & \hspace{3mm}0\hspace{3mm} & \hspace{3mm}0\hspace{3mm} &
   \hspace{3mm}0\hspace{3mm} & \hspace{3mm}0\hspace{3mm} & \hspace{3mm}0\hspace{3mm} & \hspace{3mm}0\hspace{3mm} \\
  0 & -\alpha & 0 & 0 & 0 & 0 & 0 & 0 \\
  0 & 0 &  \beta & 0 & 0 & 0 & 0 & 0 \\
  0 & 0 & 0 & \frac13 \beta & 0 & 0 & i\frac{2\sqrt2}{\sqrt3} \beta & 0 \\
  0 & 0 & 0 & 0 & -\frac13 \beta & 0 & 0 & -i\frac{2\sqrt2}{\sqrt3} \beta \\
  0 & 0 & 0 & 0 & 0 & -\beta & 0 & 0 \\
  0 & 0 & 0 & -i\frac{2\sqrt2}{\sqrt3} \beta & 0 & 0 & -\frac13 \beta & 0 \\
  0 & 0 & 0 & 0 & i\frac{2\sqrt2}{\sqrt3} \beta & 0 & 0 & \frac13 \beta
  \end{array}\right),
\end{equation}
\end{widetext}
where the $z$-axis is chosen in the direction of the magnetization and $N_0 \alpha$ and $N_0 \beta$ are the $s$-$d$ and $p$-$d$ exchange constants.

The mean field value of localized spins is determined as the thermodynamical average
\begin{equation}\label{Sz}
  \langle S \rangle = \langle \hat{S}_z \rangle = \frac1{Z} \; {\rm Tr} \; e^{-\frac{ \hat{H}_m}{k T} } \hat{S_z},
\end{equation}
with the partition function
\begin{equation}
  Z={\rm Tr} \; e^{-\frac{\hat{H}_m}{k T}}.
\end{equation}
Within the mean field approximation for uncorrelated spins the spin Hamiltonian is
\begin{equation}
\hat{H}_m= - B_{\rm eff} \hat{S}_z,
\end{equation}
with the effective field
\begin{equation}\label{Beff}
B_{\rm eff}= \langle \hat{S}_z \rangle J_0,
\end{equation}
 and
\begin{equation}
J_0=\frac{3 k T_c}{S(S+1)}.
\end{equation}
The Curie temperature $T_c$ is an input parameter of our model; through the transcendental equations
(\ref{Sz}) and (\ref{Beff}) it determines the mean field value of $\langle S \rangle$.

\section{Local field factors for partially spin polarized systems}\label{app:lff}
Expressions for local field factors of partially spin polarized electron gas were derived in
Ref.~\onlinecite{UllrichFlatte02}, but in a different spin basis. Here we will briefly rederive them in the basis of Eq.~(\ref{rho4}).

In the adiabatic approximation (which ignores frequency dependence), the components of the tensor $\uuline{f_{\xc}}$ of the
local field factors in Eq.~(\ref{tddft}) have the form
\begin{equation}\label{fij}
  f_{ij}=\frac{\de^2\left[ n e_{\xc}(n,\xi)\right]}{\de \rho_i \de \rho_j},
\end{equation}
where $e_{\xc}$ is the  xc energy per particle, $n\equiv \rho_1$ is the electron density, and $\xi$ is the spin polarization:
\begin{equation}
  \xi\equiv |\vec{\xi}|=\frac1{n}\sqrt{\rho_z^2+\frac12(\rho_+\rho_-+\rho_-\rho_+)}.
\end{equation}
We assume here that $e_{\xc}$ depends only on the absolute value of $|\xi|$.
Direct evaluation of Eq.~(\ref{fij}) gives
\begin{eqnarray*}
  f_{11}&=& 2\frac{\de e_{\rm xc}}{\de \rho_1}
  -2\xi\frac{\de^2 e_{\rm xc}}{\de \rho_1 \de\xi}+\rho_1 \frac{\de^2 e_{\rm xc}}{\de \rho^2_1}
  +\frac{\xi^2}{\rho_1}\frac{\de^2 e_{\rm xc}}{\de \xi^2},   \\
  f_{1i}&=& \frac{\de \xi}{\de \rho_i}
  \left(\rho_1\frac{\de^2 e_{\rm xc}}{\de \rho_1 \de\xi}-\xi\frac{\de^2 e_{\rm xc}}{\de \xi^2} \right), \;\; i=(z,+,-),\\
  f_{zz}&=& A+\rho_z^2 B, \\
  f_{z+}&=& \frac{\rho_z \rho_-}2 B, \\
  f_{z-}&=& \frac{\rho_z \rho_+}2 B, \\
  f_{++}&=&\frac{\rho_- \rho_-}4 B, \\
  f_{--}&=&\frac{\rho_+ \rho_+}4 B, \\
  f_{+-}&=& \frac{A}2+\frac{\rho_- \rho_+}4 B,
\end{eqnarray*}
with
\[ A=\frac1{\rho_1 \xi}\frac{\de e_{\rm xc}}{\de \xi},\;\;\;\;\;\;\;\;\;\;\;
B=\frac1{(\rho_1 \xi)^2}\left(\xi\frac{\de^2 e_{\rm xc}}{\de\xi^2}-\frac{\de e_{\rm xc}}{\de \xi}\right).\]
Note that $f_{ii'}=f_{i'i}$ and, generally, the tensor of local field factors is a symmetric matrix. If, however,
the $z$-axis is directed along the average spin direction, so that the ground-state transverse spin densities vanish
($\rho_+=\rho_-=0$), then the matrix reduces to the block-diagonal form of Eq.~(\ref{fxc}).

We define the xc energy of the spin polarized system in the usual manner as \cite{UllrichFlatte02}
\begin{equation}
  e_{\rm xc}(n,\xi)=e_{\rm xc}(n,0)+\Big(e_{\rm xc}(n,1)-e_{\rm xc}(n,0)\Big)f(\xi),
\end{equation}
with
\begin{equation}
  f(\xi)=\frac{(1+\xi)^{4/3}+(1-\xi)^{4/3}-2}{2(2^{1/3}-1)}.
\end{equation}
This is exact for the exchange part, but only approximately so for the correlation part (which will be neglected anyway in the following).
With this, we get
\begin{equation}
  \frac{\de e_{\rm xc}}{\de \xi}=\Big(e_{\rm xc}(n,1)-e_{\rm xc}(n,0)\Big)\frac{(1+\xi)^{1/3}-(1-\xi)^{1/3}}{\frac32(2^{1/3}-1)},
\end{equation}
and
\begin{equation}
  \frac{\de^2 e_{\rm xc}}{\de \xi^2}=
  \Big(e_{\rm xc}(n,1)-e_{\rm xc}(n,0)\Big)\frac{(1+\xi)^{-2/3}+(1-\xi)^{-2/3}}{\frac92(2^{1/3}-1)}.
\end{equation}
This completes the definition of the local field factors for partially spin polarized system. The only
remaining ingredients we need to perform the actual calculations are the expressions for the xc
energy for unpolarized and fully spin polarized  system, $e_{\rm xc}(n,0)$ and $e_{\rm xc}(n,1)$.
In this work for simplicity we limit ourselves to the exchange part of $e_{\rm xc}$:
\begin{eqnarray}
  e_{\rm x}(n,0)&=& -\frac{3e^2}{4K} \left(\frac{3 n}{\pi}\right)^{1/3},
  \\
  e_{\rm x}(n,1)&=& 2^{1/3} e_{\rm x}(n,0),
\end{eqnarray}
where $K$ is the static dielectric constant of the host material.
Direct evaluation gives the following expressions:
\begin{eqnarray*}
  \frac{\de e_{\rm xc}}{\de n} &=&-\frac{e^2}{8K}\left(\frac3{\pi}\right)^{1/3} n^{-2/3}
  \left((1+\xi)^{4/3}+(1-\xi)^{4/3}\right), \\
  \frac{\de^2 e_{\rm xc}}{\de n^2} &=& \frac{e^2}{12K} \left(\frac3{\pi}\right)^{1/3} n^{-5/3}
  \left((1+\xi)^{4/3}+(1-\xi)^{4/3}\right), \\
 \frac{\de^2 e_{\rm xc}}{\de n \, \de \xi} &=& -\frac{e^2}{6K} \left(\frac3{\pi}\right)^{1/3} n^{-2/3}
  \left((1+\xi)^{1/3}-(1-\xi)^{1/3}\right).
\end{eqnarray*}

\end{document}